\documentclass{iucr}              

     \journalcode{S}              

\begin{document}                  



\title{Electron and photon diagnostics for plasma acceleration based FELs}


\cauthor[a]{Labat}{Marie}{marie.labat@synchrotron-soleil.fr}{address if different from \aff}
\author[a]{El Ajjouri}{Moussa}
\author[a]{Hubert}{Nicolas}
\author[a]{Andre}{Thomas}
\author[a]{Loulergue}{Alexandre}
\author[a]{Couprie}{Marie-Emmanuelle}

\aff[a]{Synchrotron SOLEIL, Saint-Aubin, 91 191 Gif-sur-Yvette \country{France}}









\maketitle                        

\begin{synopsis}
Supply a synopsis of the paper for inclusion in the Table of Contents.
\end{synopsis}

\begin{abstract}
It is now well established that laser plasma acceleration (LPA) is an innovative and good candidate in the beam acceleration field. Relativistic beams are indeed produced up to several GeV but their quality remains to be demonstrated in the highly demanding case of Free Electron Lasers (FELs). Several experiments already showed the feasibility of synchrotron radiation delivery based on a LPA but free electron lasing has still to be achieved. Since the quality of the LPA beam inside the undulator is the critical issue, any LPA based FEL experiment requires a refined characterization of the beam properties along the transport line and of the photon beam at the undulator exit. This characterization relies on diagnostics which must be adapted to the LPA specificities. 
We will review the electron and photon diagnostics already used on LPAs and required for LPA based FELs, and illustrate the critical points using recent experiments performed around the world.
\end{abstract}


\section{Introduction}

Free Electron Lasers (Madey \textit{et al.}, 1971) (FELs) are presently the most brilliant light sources and precious tools for time-resolved studies of molecular and atomic dyanmics (Zewail \textit{et al.}, 2003). 
Several facilities are rountinely in operation for users in the hard X-ray range (Emma  \textit{et al.}, 2010; Ishikawa  \textit{et al.}, 2012) as in the XUV and soft X-ray range (Ackermann  \textit{et al.}., 2007; Allaria  \textit{et al.}, 2013).
An FEL is a system consisting of a relativistic electron beam and an undulator, which enables to amplify the spontaneous synchrotron radiation created by the particles oscillating in the undulator, or an external seed injected together with the electron beam in the undulator. This amplification is exponential in $e^{z/L_G}$, $z$ being the longitudinal coordinate in the undulator and $L_G$ the so-called gain length (Dattoli \textit{et al.}, 1993; Kim \textit{et al.}, 1993). This gain length is directly related to the electron beam parameter according to $L_G = \frac{\lambda_u}{4\pi \sqrt{3}} \times {\gamma} (\frac{\sigma_x \sigma_y}{\hat{I}})^{1/3}$, with $\lambda_u$ the undulator period, $\gamma$ the usual Lorentz factor, $\hat{I}$ the peak current and $\sigma_{x,y}$ the transverse beam size. 
A high gain FEL therefore corresponds to a short gain length system, which requires a high density, i.e. high quality electron beam.

Present FELs are based on Radio-Frequency Accelerators (RFAs) which offer the highest quality electron beams: typically charges of 10~pC up to 1~nC with a normalized emittance between 0.1 and 1$\pi$.mm.mrad and an energy spread which does not exceed the 0.01~$\%$ level. Thanks to decades of operation and improvments, the stability of those parameters can stay in the percent level during typically days of operation. Nevertheless, their accelerating gradient is limited to 100~MV/m.

In the late 1970s, a compact and elegant alternative to the RF technology popped-out with Tajima and Dawson (Tajima \textit{et al.}, 1979) which proposed laser generated plasma wakes to accelerate particles.
Impressive developments (Pittman \textit{et al.}, 2002; Pukhov \textit{et al.}, 2002) led in 2004 to the first demonstration of plasma acceleration (Mangles \textit{et al.}, 2004; Geddes \textit{et al.}, 2004; Faure \textit{et al.}, 2004). Laser Plasma Accelerators (LPAs) were born. Since then, the quality of LPA beams has kept increasing (Chien \textit{et al.}, 2005; Leemans \textit{et al.}, 2006; Faure \textit{et al.}, 2006; Geddes \textit{et al.}, 2008). The typical LPAs parameters, a charge of a few pC, a normalized emittance of a few $\pi$.mm.mrad, an energy spread above a few $\%$ and a divergence of a few mrad, still remain far below the RFAs' in terms of FEL requirements. But LPAs can deliver above 100~GV/m accelerating gradients, which could be one way towards more compact accelerators and, consequently, towards more compact FELs. Together with the novelty of the technology, this motivated several groups to try the operation of an LPA based FEL.

In 2008, Schlenvoigt at al. (Schlenvoigt \textit{et al.}, 2008) observed the first Synchrotron Radiation (SR) resulting from an LPA beam travelling through an undulator. Using a 60~MeV beam together with a 1~m long undulator of 20~mm period, they recorded an SR spectrum centered at 740~nm. Four other groups then also successfully produced SR both in the UV and visible range (Fuchs \textit{et al.}, 2009; Lambert \textit{et al.}, 2012; Anania \textit{et al.}, 2014; Couprie \textit{et al.}, 2017). The key features of these experiments are summarized in Table~\ref{table:exp} and the COXINEL most advanced layout is given as example in Figure 1. At least two other groups are presently working on the same topic: Leemans \textit{et al.} at Lawrence Berkeley National
Laboratory in the U.S. and Sano \textit{et al.} on the ImPACT project in Japan.

But none of the experiments which produced SR, succeded yet into amplifying this SR. The gain resulting from the large divergence and energy spread of the LPA beam remains to low with respect to the undulator length used.

To deal with those realistic high energy spread and high divergence, three techniques have been proposed: horizontally disperse and couple the beam in a Transverse Gradient Undulator (Huang \textit{et al.}, 2012), decompress and couple the beam in a suitably tapered undulator (Maier \textit{et al.}, 2012; Seggebrock \textit{et al.}, 2013; Couprie \textit{et al.}, 2014) and finally decompress and synchronize the beam waist advance with the FEL slippage (Loulergue \textit{et al.}, 2015). While previous methods have never been implemented, the chromatic matching is presently being tested on COXINEL (Couprie \textit{et al.}, 2016).

Whatever the final configuration, the demonstration and further operation of an LPA based FEL requires dedicated diagnostics which should be adapted to the LPAs specificities. In this paper, we review the main diagnostics operated on LPAs for electron beam characterization and manipulation in transport lines, adressing main issues and recent achievements. We also review the photon diagnostics used up to now in the attempts of LPA based FEL experiments, underlying the main difficulties related to their implement on an LPA line.

\section{Electron beam diagnostics}
\subsection{Charge measurements}

Charge measurements on RFAs mainly rely on Faraday cups (Brown \textit{et al.}, 1956), Integrating Current Transformers (ICTs) (Unser \textit{et al.}, 1989) and Rogowski coils (Rogowski \textit{et al.}, 1912). But the very strong Electro-Magnetic Pulse (EMP) environment of the LPAs tends to spoil any electronic based measurements. This is why charge measurements on LPAs essentially rely on photoluminescence based detectors, among which the most popular are Imaging Plates (IPs) and Lanex screens.

\subsubsection{Imaging plates (IPs)}

An IP is a multilayer film with a photostimulable phosphor layer. Irradiation by an electron beam excites electron-hole pairs in the sensitive phosphor, which can remain trapped and later detrapped for detection using photostimulated luminescence with a scanner. IPs have been calibrated on RFAs versus beam energy up to 100~MeV using Rogowski coils (Tanaka  \textit{et al.}, 2005) and versus charge up to 120~pC using ICTs (Zeil \textit{et al.}, 2010).
Their dynamic range is large ($\approx 10^5$) and they offer a high sensitivity. But before processing the films, one has to wait for the electron-hole pairs decay time to stabilize in time, which typically requires a couple of hours. In spite of this "fading" effect which makes them unsuitable for high repetition rate measurements, they remain a reference in terms of absolute charge measurement in the community (Tanaka \textit{et al.}, 2005).

\subsubsection{Lanex screens}

Lanex screens are scintillators consisting of mixture of phosphor powder in a urethane binder. When the electron beam passes through the screen, it deposits energy which results into light emission in the visible range. The photon distribution can then simply be imaged by an objective on a Charge Couple Device (CCD) camera. Lanex screens have been calibrated on RFAs up to an energy of 1.5~GeV and for charges up to 800~pC (Nakamura \textit{et al.}, 2011). Since then, they are extensively used on LPAs to provide absolute charge measurements. It is this technique which has been used on all but COXINEL LPA based SR experiments to measure bunch charges in the 0.1 to 30~pC range (Schlenvoigt \textit{et al.}, 2008; Fuchs \textit{et al.}, 2009; Lambert \textit{et al.}, 2012; Anania \textit{et al.}, 2014).

\subsubsection{ICTs}

Whereas IPs and Lanex screens are destructive measurements, ICTs are non-invasive. But relying on an electronic which integrates the charge over a time window of hundreds of $\mu$s, they are perfectly adapted to the LPAs' strong EMP environment, low charge beams and single-shot requirement. Nevertheless, ICTs from BERGOZ were implemented on LPAs and compared to Lanex screen measurements.
In a first work (Glinec \textit{et al.}, 2006), a descrepancy up to a factor 8, varying shot-to-shot, was reported. But in a more recent work (Nakamura \textit{et al.}, 2011), a very good agreement was obtained. The (new) reliability of the ICT could be attributed by the authors to :
(i) a special care taken to avoid EMP effects (cable extension for time separation of signals, cable shielding, arranged route), (ii) the installation of a metallic foil and of a low acceptance aperture to prevent particle/radiation hit on the ICT, (iii) the installation out of vaccum and on a ceramic gap of the ICT so that e-beams propagate in vacuum with minimum disturbance and (iv) the large spacing of ICT and Lanex with respect to the source point (4~m) to prevent low energy electrons to reach the ICT coil. 

Since then, a new generation of ICTs, turbo-ICTs, was developed (BERGOZ). Relying on the same physical principle as ICTs, they are coupled to a low noise amplifier and an RF modulator which makes them optimized for low charges (down to 10~fC) measurements. In addition, operated in the single bunch mode, they can detect sub-ns bunch charges of less than 1~pC charge. Turbo-ICTs have been implemented on an LPA for the first time on the COXINEL experiment (Labat \textit{et al.}, 2014; Couprie \textit{et al.}, 2016). One item was placed at the LPA source exit (i.e. inside the electron beam generation vacuum chamber) and another at the exit of the undulator. In both cases, a Lanex screen monitor was implemented just downstream the turbo-ICT for bunch charge comparative measurements. As expected, the first turbo-ICT suffered from the strong EMP in the vicinity of the e- beam source and provided with incoherent data. But the second one, 10~meters downstream, gave charge measurements in very good agreement with the non-absolute charge measurements (in CCD counts units) of the downstream Lanex screen. The correlation over 30 consecutive shots is illustrated in Figure~2.

\subsubsection{Spectrum measurement techniques}

On RFAs, electron beam spectrum measurements commonly rely on a dispersive element which spatially spans the beam on a detector. 
The same technique can be used on LPAs, provided that: (i) due to the very broad energy range expected, the detection area is large enough and (ii) that to deal with the high shot-to-shot fluctuations, the full spectrum can be acquired in one single-shot.
Variable magnetic fields in front of a fixed silicon detector were first used (Fritzler \textit{et al.}, 2004; Malka \textit{et al.}, 2001) but lacked the single-shot specification. Compact dipoles together with IPs (Tanaka \textit{et al.}, 2005; Clayton \textit{et al.}, 1995) were also tried but were limited in terms of repetition rate. Finally, after scintillating fiber arrays (Sears \textit{et al.}, 2010), scintillating screens appeared as the most adapted detector. Coupled to an appropriate imaging system, they enable -in one single-shot- full range spectrum, charge and divergence measurement (Glinec \textit{et al.}, 2006).
A suitable choice of dipole strength, screen size and imaging magnification enables to reach the desired resolution without fundamental limit.

\subsubsection{Spectra on LPA based SR experiments}

In all LPA based SR experiments (Schlenvoigt \textit{et al.}, 2008; Fuchs \textit{et al.}, 2009; Lambert \textit{et al.}, 2012; Anania \textit{et al.}, 2014; Couprie \textit{et al.}, 2016), the electron beam spectrum was recorded using this technique.
Even if the screen type (Konica TR in (Schlenvoigt \textit{et al.}, 2008), phosphor screen in (Fuchs \textit{et al.}, 2009), Lanex and Ce:YAG in (Anania \textit{et al.}, 2014; Couprie \textit{et al.}, 2016), or the dipole type (permanent magnet in (Schlenvoigt \textit{et al.}, 2008; Fuchs \textit{et al.}, 2009; Anania \textit{et al.}, 2014) or electro-magnetic in (Couprie \textit{et al.}, 2016) change from one setup to the other, all spectra are recorded in single-shot and enable to exhibit a broad-band energy distribution with a fluctuating mean value as well as spreading. An illustrative example is presented in the bottom-half of Figure~3.

\subsection{Position measurements}

\subsubsection{Position measurement techniques}

For an accurate alignment and focussing of the electron beam inside the undulator, mandatory for SR and further FEL light production, the electron beam position must be precisley measured.
The standard instruments on RFAs (in particular LINACs) are striplines (Suwada \textit{et al.}, 2000) and cavity BPMs (Hartman \textit{et al.}, (1995); Keil \textit{et al.}, 2010; Keil \textit{et al.}, 2013). Non-destructive, easily included into a feed-back system, they provide with a resolution which can reach the sub-micron level. 

But for LPAs, because of the large beam pointing (mrad) and shape fluctuations, the devices should allow a large detection area and be as insensitive as possible to the bunch profile. This is why up to now, a scintillating screen imaged on a CCD has been the preferred method. The final resolution can raise up to 500~$\mu$m and is hardly below the 10~$\mu$m level depending on the scintillator type, but it is most of the time enough due to the large divergence of the beams.

\subsubsection{Position on LPA based SR experiments}

All LPA based SR experiments used screen monitors for position measurements. 
They were first used for a rough observation of the beam position and shape, i.e. to check that the beam came more or less in and out of the undulator (Schlenvoigt \textit{et al.}, 2008; Fuchs \textit{et al.}, 2009; Lambert \textit{et al.}, 2012).
The first characterization of an LPA beam along a transport line was presented in (Anania \textit{et al.}, 2014). Using four screen monitors distributed along the line, the beam shape evolution was observed and a beginning of a matching was tried in the undulator. Still, there was no real control of the beam phase-space through out the line. This was first achieved on COXINEL. Using six screen monitors, the beam could be observed: (i) at the LPA source exit, with or without the first triplet of quadrupoles to finally adjust the first stage of refocussing, (ii) in the middle of the chicane, with a none zero dispersion, to measure the beam energy and energy spread, (iii) and (iv) at the undulator entrance and exit to finely tune the beam matching inside the undulator and (vi) after the final dump dipole to remeasure the beam energy distribution at the end of the line. 
The beam manipulation mastering could essentially be performed thanks to the use of Ce:YAG screens instead of Lanex screens at the entrance and exit of the undulator (Labat \textit{et al.}, 2014). In Figure~4, is presented a series of 10 consecutive shots recorded on the screen monitor located at the undulator exit. For the first 5 shots a Lanex screen was used while for the last 5 a Ce:YAG screen was inserted (using a motorized stage to flip from one to the other). The Ce:YAG, thanks to its intrinsec higher resolution, enables to distinguish -thus with a lower photon yield- much more refined structures. Comparing the observed beam shape to the one expected from tracking simulations, we could finely optimize the transport down to the undulator.

The first installation of cavity BPMs (cBPMs) on an LPA was also realized on COXINEL (Labat \textit{et al.}, 2014; Couprie \textit{et al.}, 2016), allowing the first on-line position measurements. A first item was placed at the undulator entrance and a second at the undulator exit, in both cases just upstream a screen monitor for comparative measurements of the beam position. This comparison was achieved using a broad-band electron beam energy i.e. spanning from 50 up to 220~MeV at the source point and from 150 to 190~MeV after spectral filtering in the chicane using an Aluminium slit.
As illustrated in Figure~5, the agreement is still not very satisfactory. The descrepancy in the absolute amplitude of the beam displacements (obtained using steerers) can reach a factor 4 and varies depending on the plane and the cBPM considered. Because of the large beam position fluctuations at the cBPM location (+/-0.5~mm while a few tens of microns stability would have been required) and low repetition rate (0.1~Hz or less), we could not achieve a proper calibration of the cBPMs. In addition, for a given machine setting, i.e. ideally fixed beam position, the cBPM gave position fluctuations far above the screen monitor ones and in an uncorrelated way. This is essentially due to the sensitivity of the cBPM to the bunch shape. cBPMs detect the center of mass of the particle distribution, while the screen monitor enables to follow the center of the high density core beam, the part of interest. 

Nevertheless, the cBPMs provided with an average position measurement in rather good relative agreement with the screen monitors, which is already one step towards on-line position measurements on LPAs. By filtering the beam in energy, we hope to improve the cBPMs' accuracy.

\subsection{Bunch length measurements}
\subsubsection{Bunch length measurement techniques}

Several techniques are in operation on RFAs. The Streak Camera (Lumpkin \textit{et al.}, 1999) is easy to implement, but limited to ps-resolutions. The Transverse Deflecting Cavities (TDS) (Loew \textit{et al.}, 1965; Behrens \textit{et al.}, 2014) can be fs-resolution but require the implement of an RF cavity which can be an issue in a non RF environment as it is the case of most LPAs.  On the other hand, Coherent Transition Radiation (CTR) analysis (Wesch \textit{et al.}, 2011;  Maxwell \textit{et al.}, 2013) and Electro-Optic Sampling (EOS) (Yan \textit{et al.}, 2000; Shan \textit{et al.}, 2000; Wilke \textit{et al.}, 2002), in their spectral or spatial encoding versions, enable single-shot and sub-ps resolution measurements. Because LPAs bunch length is typically of the order of the plasma wavelength, i.e. a few $\mu$m, fs-resolution is required, and again to cope with the large fluctuations, the measurement should be single-shot. CTR and EOS were therefore naturally the methods implemented on LPAs (not yet on LPA based SR experiments).

\subsubsection{Bunch length measurements on LPAs}

In 2006, the EOS technique in the temporal encoding version was used to measure bunch lengths of the order of 50~fs-rms (Van Tilborg \textit{et al.}, 2006).
The CTR technique first enabled to evidence a sub-micron modulation of LPA beams at the laser wavelength in the laser plane of polarization, in good agreement with PIC simulations (Glinec \textit{et al.}, 2007). A few years later, CTR was used to measure LPA bunch lengths of the order of 1.4-1.8~fs-rms (Lundh \textit{et al.}, 2011).

\subsection{Emittance measurements}
\subsubsection{Emittance measurement techniques}

The emittance is well known as the figure of merit for relativistic particles since it quantifies its divergence and focusability. But is also a key parameter of the FEL gain and final brightness. Several methods have been proposed and implemented on RFAs to measure the geometric emittance. The quadrupole scan (Minty \textit{et al.}, 2003) is a reliable thus not single-shot possibility. 
Using multiscreen image analysis at different betatron phases (Cutler \textit{et al.}, 1987; Yakimenko \textit{et al.}, 2002) is neither single-shot and in addition requires a long and complicated transport, little suitable for LPAs. The multiple OTR screen analysis (Thomas \textit{et al.}, 2011) was demonstrated to be another single-shot technique, though for GeV range beam energies. Finally the pepper-pot technique (Zhang \textit{et al.}, 1998; Yamazaki \textit{et al.}, 1992), probing the phase-space with a mask of holes or slits, may be the only single-shot method for $<$ GeV beams.

\subsubsection{Emittance measurements on LPAs}

The first pepper-pot 2D and single-shot emittance measurement on an LPA beam (Brunetti \textit{et al.}, 2010) gave a normalized emittance of $< \epsilon_{nx} >$ = 2.2 +/-0.7~$\pi$.mm.mrad in the horizontal and $< \epsilon_{nz} >$ = 2.3+/-0.6~$\pi$.mm.mrad in the vertical plane at 125~MeV (see Figure~6). But a few years later, the limitations of this technique in the case of LPA beams were clearly adressed (Cianchi \textit{et al.}, 2013). LPA beams exhibit an ultra-thin phase-space due to their large divergence. The pepper-pot method proposes to sample this phase-space using holes or slits. Whatever the mask, the high thickness of the phase-space may lead to an inefficient sampling resulting in large errors on the emittance estimate. In the case reported in (Brunetti \textit{et al.}, 2010), this error might reach 47~$\%$ assuming a 10~$\mu$m initial spot size and even 1000~$\%$ assuming a 1~$\mu$m initial spot size. The Quadrupole scan method, on the other hand, in spite of being single-shot, might only be limited by the beam size inside the quadrupoles. Provided this size is small enough to avoid chromatic effects and consequent emittance dilution, the method should remain reliable in the case of LPAs.
This method was actually implemented with sucess to measure sub mm.mrad emittance as reported in (Weingartner \textit{et al.}, 2012). In the same publication, an alternative single-shot method was also proposed and demonstrated: the energy scan. Using two quadrupoles to focuss the beam in the horizontal plane and one dipole to disperse the beam in energy in the vertical plane, a 2D beam distribution can be recorded on a scintillator. Further fitting the horizontal beam size as a function of the beam energy, provides the geometric emittance in one single-shot (see Figure~7). Both methods, quadrupole and energy scan were found in 10~$\%$ agreement.
Finally, it is also possible to estimate an LPA beam emittance using spectroscopy (Plateau \textit{et al.}, 2012) : it is a single-shot thus indirect measurement.

\section{Photon Diagnostics}
\subsection{Background issues}
Prior implement of photon diagnostics, background issues should be adressed. Indeed, LPAs suffer from a very "high-light" environment resulting from at least three type of sources. The IR laser which is used for electron beam generation is tightly focussed at the source point and therefore very divergent downstream. Nevertheless, because it is also ultra-intense and unfortunatly well guided by metallic vacuum pipes, the laser intensity remains nothing but negligible even meters downstream. In all LPA based SR experiments, aluminium foils on the laser path were used to cut this IR laser. But, though not estimated in the corresponding publications, the effect of those foils can be dramatic on the slice emittance. Applying simple analytical formula (Chao \textit{at al.}, 1999) to a typical LPA beam (100~MeV, 1 $\pi$.mm.mrad slice emittance) with a magnification of 1/20 in the undulator, we found that an Aluminium foil on the beam path would multiply the slice emittance by a factor 10 for a thickness of 15~$\mu$m as used in (Schlenvoigt \textit{et al.}, 2008; Fuchs \textit{et al.}, 2009) and by a factor 2 for a thickness of 500~nm as more or less used in (Anania \textit{et al.}, 2014; Couprie \textit{et al.}, 2017). 
The only alternative to our knowledge would be the use of a dogleg, but this would lead to other issues of transport.

The plasma created by the intense IR laser also produces a strong isotropic illumination in the visible range, together with a wide range of particles (X-rays, Gamma-rays, etc.). This parasitic "light" can easily reach all the detection system implemented in the accelerator room and requires all the diagnostics, and in particular photon diagnostics, to be carefully protected using bandpass filters, blockers, shielding, etc..

In addition to these, comes the coherent and incoherent radiation which is systematically emitted when the electron beam crosses the plasma-vacuum transition at the source exit and the metallic foil previously mentionned. This component may often be negigible with respect to the previous ones, but should still be adressed.

\subsection{Spectrum measurements}
The present LPA based SR experiments and close future LPA based FEL experiment do not have to deal with users and can afford a destructive spectrum measurement. Therefore, basic spectrometers relying on a grating and a CCD were simply implemented, eventually coupled to a collecting optics. Because the initial LPA beam highly fluctuates in energy, the resulting SR spectra are also expected to fluctuate, requiring both single-shot photon measurements and simultaneous record of the electron beam spectral content.
The most illustrative example of LPA based SR spectrum measurement is presented in Figure~3. It first shows that this kind of spectrometer (grating with CCD) enables the measurement of both the beam energy content (along the horizontal axis) and divergence (along the vertical axis). It also shows that the central SR wavelength can fluctuate by more than 25~$\%$ because of the initial beam energy variations.

\subsection{Beam profile measurements}
Like on synchrotron beamlines or FEL beamlines, the radiation profile can be measured directly using a CCD or indirectly using an intermediate scintillator. 

The first beam profile measurement of LPA based SR was reported in (Lambert \textit{et al.}, 2012) and is shown in Figure~8. The footprint was recorded without any spectral filtering, i.e. in a wide (230-440~nm) range, and the SRW (Chubar \textit{et al.}, 1998) simulations only matched in one plane, probably because of remaining parasitic light in the horizontal direction. Beam profiles systematic measurements were recently achieved on COXINEL. But their analysis is still on-going.

\section{Conclusion}

Of course, diagnostics for LPA beams and LPA based SR are not as advanced as those developped for RFAs. But LPA is a much more recent technology which still needs time to adapt or develop its diagnostics. 
The basic tools already exist for a full characterization of an LPA beam along a complex transport line, and for the analysis of consequent SR or FEL radiation. 
But both beam characterization and further FEL applications would highly benefit further improvment of on-line diagnostics to survey the LPA stage.

The demonstration of an LPA based FEL remains to be done, but thanks to advanced ideas of beam manipulation, this no longer stands in a far future. 

\ack{Acknowledgements}

The authors wish to thank the European Research Council for COXINEL Grant (340015), together with the COXINEL team at SOLEIL, the LOA group of V. Malka, the Diagnostics and the Synchronization Group of Synchrotron SOLEIL.


\begin{table}
\caption{\label{table:exp}LPA based SR experiments which reported SR observation at an undulator exit. $E_e$: electron beam energy in MeV, $\lambda_u$: undulator period in mm, $L_u$ undulator length in m, $\lambda_r$: resonance wavelength in nm. Trans. for transport magnetic elements. Q-lenses: magnetic lenses, EMQ: electro-magnetic quadrupole, PMQ: permanent magnet quadrupole.}
\begin{tabular}{lccccc}      
 Exp.    	& IOQ Jena     	& MPQI       	& LOA     	& Univ. of 				& LOA and	\\
 			& 				& 				& 			& Strathclyde 			& SOLEIL \\
 Country 	& Germany		& Germany 		& France 	& UK 					& France \\
\hline
 $E_e$      	& 60    & 200      	& 100-150	& 100	& 176   \\
 $\lambda_u$  	& 20    & 5      	& 18.2    	& 15 	& 18 	\\
 $L_u$      	& 1     & 0.3      	& 0.6      	& 0.5	& 2 	\\
 $\lambda_R$ 	& 740 	& 18 		& 250-400 	& 220 	& 200	\\
 Transp. 		& - 	& 2 Q-lenses 	& 3 EMQs	& 3 PMQ 	& 3 PMQ		\\
 				& 		& 			& 			& + 	3 EMQs 	& + 4 dipoles \\
 				& 		& 			& 			&  				& + 4 EMQs \\
\end{tabular}
\end{table}

\begin{figure}
\includegraphics{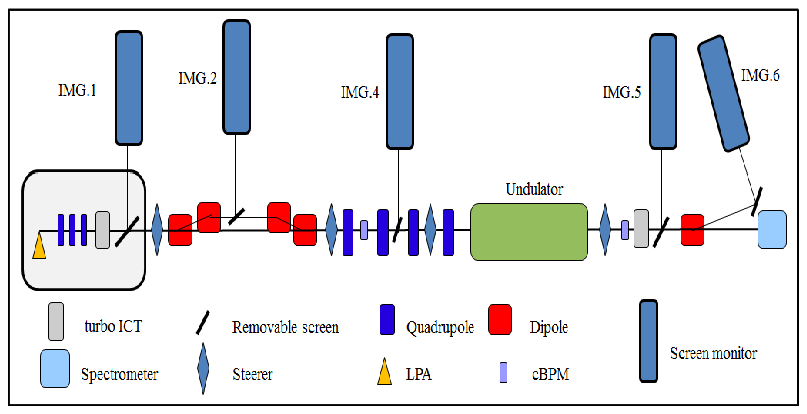}
\caption{Layout of the COXINEL experiment consisting from left to right of: an LPA (IR laser focussed into a gas jet), three Permament Magnet Quadrupoles (QUAPEVA) to refocuss the beam, a turbo ICT, a screen monitor (IMG.1), a steerer to correct the beam orbit, four dipole forming a chicane to decompress the beam with a screen monitor (IMG.2) in its middle, a second steerer, four electro-magnetic quadrupoles to match the beam in the undulator with in between first and second a cavity Beam Position Monitor (cBPM), in between second and third a fourth screen monitor (IMG.4) and in between third and fourth a third steerer, an undulator of 110 periods of 18~mm, a fourth steerer, a last cBPM, a last turbo ICT, a screen monitor (IMG.5), a dump dipole, a last screen monitor (IMG.6) and a spectrometer in the visible range.}
\end{figure}

\begin{figure}
\begin{center}
\includegraphics{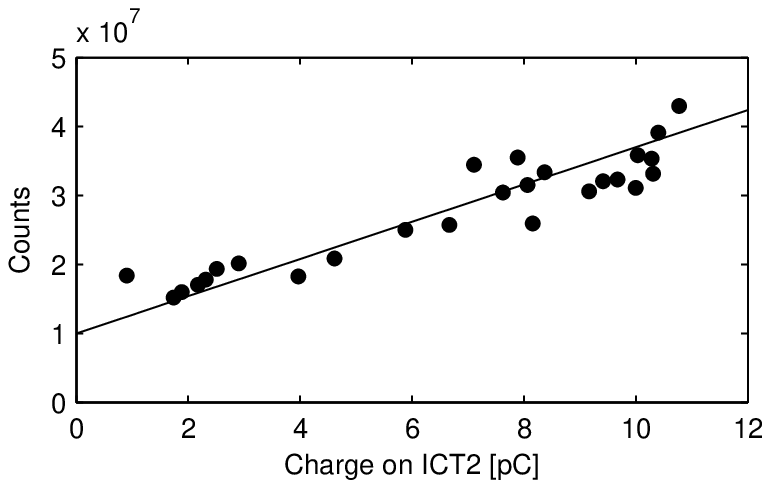}
\caption{Charge measurements on the COXINEL experiment: number of counts recorded on the screen monitor IMG.5 within a 500~pixels diameter ROI versus charge measured on the turbo-ICT at the undualtor exit. (Dots) Measurement, (--) Linear fit.}
\end{center}
\end{figure}

\begin{figure}
\includegraphics{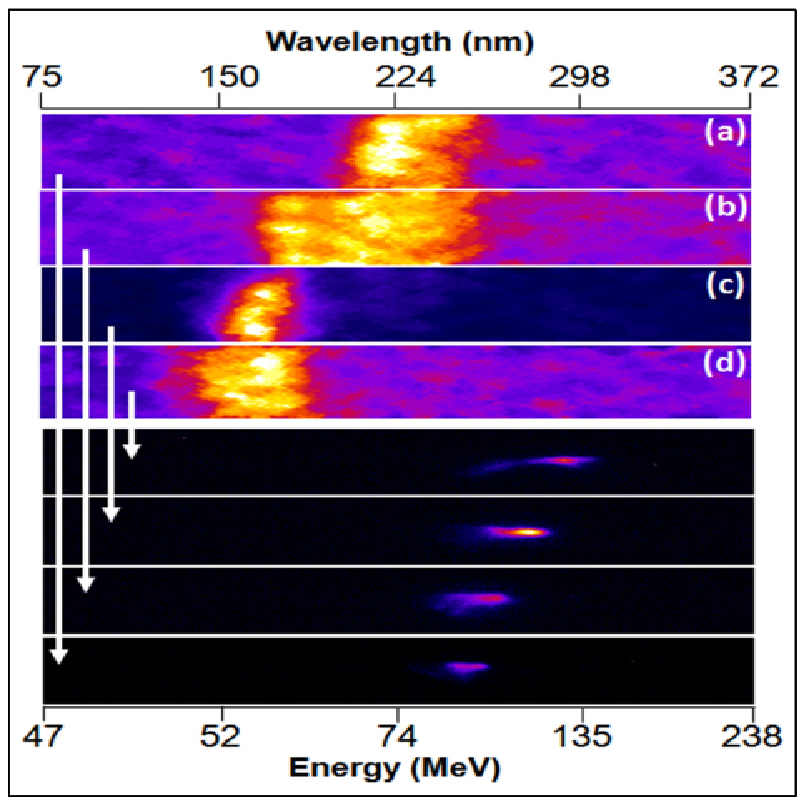}\\
\caption{False color images of four unprocessed undulator radiation spectra with corresponding electron spectra indicated. Respective values for number of detected photons (after processing for toroidal mirror, grating, and camera response), electron beam charge, and central energy are (a) 1.2 $\times 10^6$, 0.9~pC, and 92 MeV, (b) 7.7 $\times 10^6$, 1.6~pC, and 95 MeV, (c) 6.1 $\times 10^6$, 2.0~pC, and 108 MeV and (d) 4.0 $\times 10^6$, 1.3~pC, and 122 MeV. Figure 3 from (Anania \textit{et al.}, 2014).}
\end{figure}

\begin{figure}
\begin{center}
\includegraphics{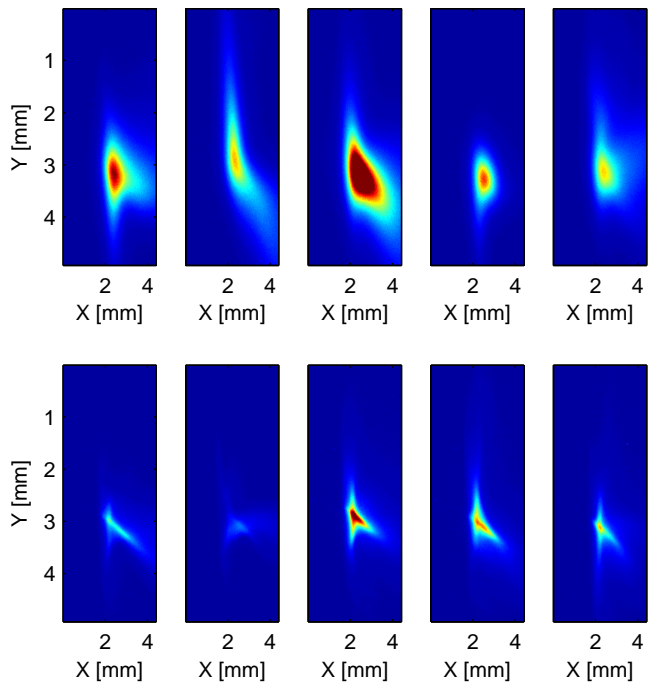}
\caption{Consecutive beam profiles recorded at the undulator exit on COXINEL experiment using (top) Lanex and (bottom) Ce:YAG screens. Color scale fixed for all images.}
\end{center}
\end{figure}

\begin{figure}
\includegraphics{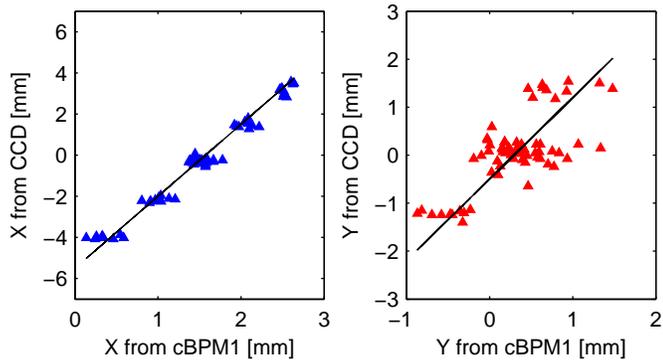}\\
(a) IMG.4 CCD versus cBPM1 measurements.\\
\includegraphics{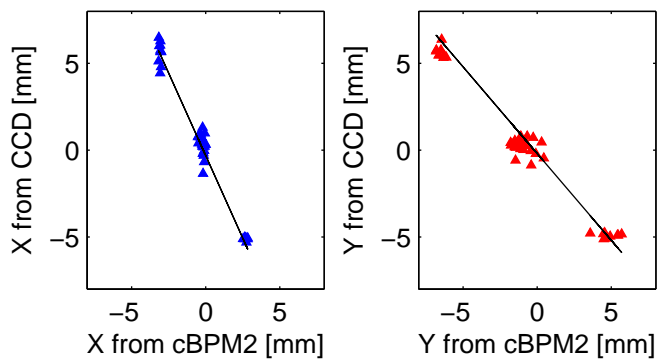}\\
(b) IMG.5 CCD versus cBPM2 measurements.\\
\caption{Ce:YAG screen versus cBPM position measurement on COXINEL (a) at the undulator entrance and (b) at the undulator exit. In both cases, the cBPM is just downstream the screen monitor.}
\end{figure}

\begin{figure}
\includegraphics{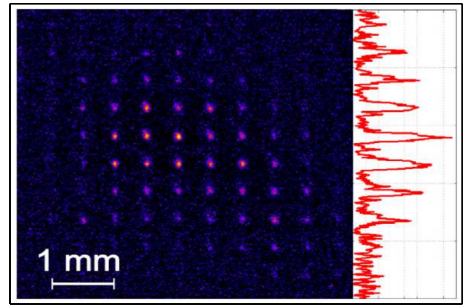}
\caption{A false color, background corrected, pepper-pot image produced on the Ce:YAG crystal by an electron beam after propagation through the emittance mask. A vertical lineout is shown on the right-hand side. Figure 2 from (Brunetti \textit{et al.}, 2010).}
\end{figure}

\begin{figure}
\includegraphics{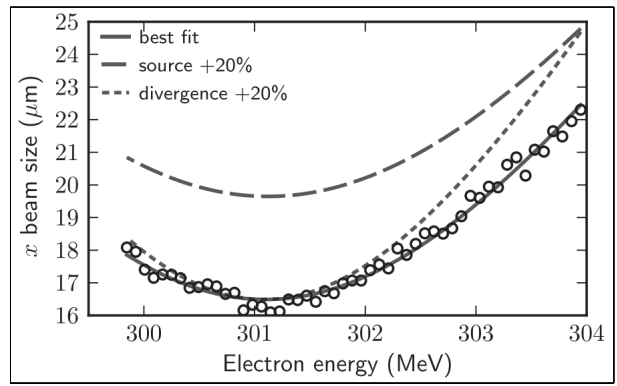}
\caption{The rms beam size vs beam energy for a single shot (circles). The solid fit line corresponds to a beam with normalized emittance of 0.14 +/- 0.01~$\pi$.mm mrad. The other lines show the expected functions for a 20~$\%$ larger emittance by varying the inferred source size or divergence. Figure 3 from (Weingartner \textit{et al.}, 2012).}
\end{figure}

\begin{figure}
\includegraphics{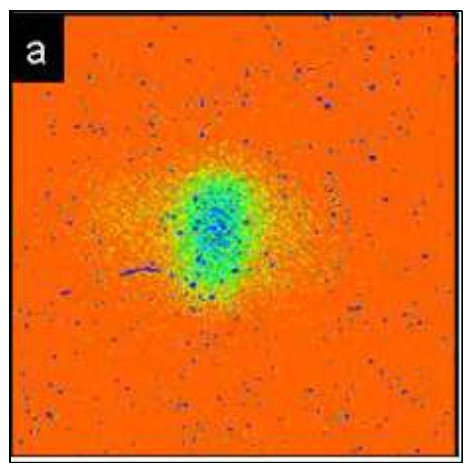}\\
\caption{Single shot footprint of the radiation measured on the CCD camera. Figure 2.a from Lambert \textit{et al.}, 2012).}
\end{figure}

\end{document}